\documentclass[aps,prl,twocolumn,groupaddress,nofootinbib,longbibliography]{revtex4-1}

\usepackage{amsmath,amsfonts,amssymb,graphicx,color,
times,bbm,psfrag,dsfont}
\usepackage{bbold}
\usepackage{mathrsfs}
\usepackage{slashed}
\usepackage{float}
\usepackage[caption=false]{subfig}
\usepackage{tikz}
\usepackage{pgfplots}
\usepackage[theorems,skins]{tcolorbox}
\newtcolorbox{mymathbox}[1][]{colback=white, ams gather, outer arc=0pt, #1}
\usepackage{mathrsfs}
\usepackage{epstopdf}
\usepackage{multirow}
\usepackage{gensymb} 
\usepackage{ulem} 
\usepackage{xr}
\usepackage{zref-xr}
\usepackage{chngcntr}
\usepackage{siunitx}
\usepackage{mathtools} 

\definecolor{myblue1}{rgb}{.14, 0.42, .84}
\definecolor{myblue2}{rgb}{0.08,0.28,0.85}
\definecolor{myred}{rgb}{0.6,0.1,0.2}
\usepackage[colorlinks=true, citecolor=myblue2, linkcolor= myred, urlcolor=myblue2,bookmarks=true]{hyperref}

\def \beq{\begin{equation}}
\def \eeq{\end{equation}}
\def \bse{\begin{subequations}}
\def \ese{\end{subequations}}
\def \bea{\begin{eqnarray}}
\def \eea{\end{eqnarray}}
\def \bem{\begin{displaymath}}
\def \eem{\end{displaymath}}
\def \bem{\begin{pmatrix}}
\def \eem{\end{pmatrix}}
\def \bs{\boldsymbol}
\def \nn{\nonumber}

\begin{document}

\title{Doped Twisted Bilayer Graphene near Magic Angles:\\ Proximity to Wigner Crystallization not Mott Insulation}
\author{Bikash Padhi}
\thanks{\href{bpadhi2@illinois.edu}{bpadhi2@illinois.edu}}
\affiliation{Department of Physics, University of Illinois at Urbana-Champaign, Urbana, Illinois, USA}
\author{Chandan Setty}
\affiliation{Department of Physics, University of Illinois at Urbana-Champaign, Urbana, Illinois, USA}
\author{Philip W. Phillips}
\thanks{\href{dimer@illinois.edu}{dimer@illinois.edu}}
\affiliation{Department of Physics, University of Illinois at Urbana-Champaign, Urbana, Illinois, USA}

\begin{abstract}
We devise a model to explain why twisted bi-layer graphene (TBLG) exhibits insulating behavior when $\nu=2,3$  charges occupy a unit moir\'e cell, a feature attributed to Mottness~\cite{cao2018Mott,cao2018SC,balents,balents2,balents3,balents4,baskar}, but not for $\nu=1$, clearly inconsistent with Mott insulation. We compute  $r_s=E_U/E_K$, where $E_U$ and $E_K$ are the potential and kinetic energies, respectively, and show that (i) the Mott criterion lies at a density $10^4$ larger than in the experiments and (ii) a transition to a series of Wigner crystalline states exists as a function of $\nu$. We find, for $\nu=1$, $r_s$ fails to cross the threshold ($r_s = 37$) for the triangular lattice and metallic transport ensues.  However, for $\nu=2$ and $\nu=3$, the thresholds, $r_s=22$, and $r_s=17$, respectively are satisfied for a transition to Wigner crystals (WCs) with a honeycomb ($\nu=2$) and kagome ($\nu=3$) structure.  We believe, such crystalline states form the correct starting point for analyzing superconductivity.
\end{abstract}

\maketitle

\begin{figure}[h]
\subfloat[]{\includegraphics[width=0.32\columnwidth , height= 0.23\columnwidth]{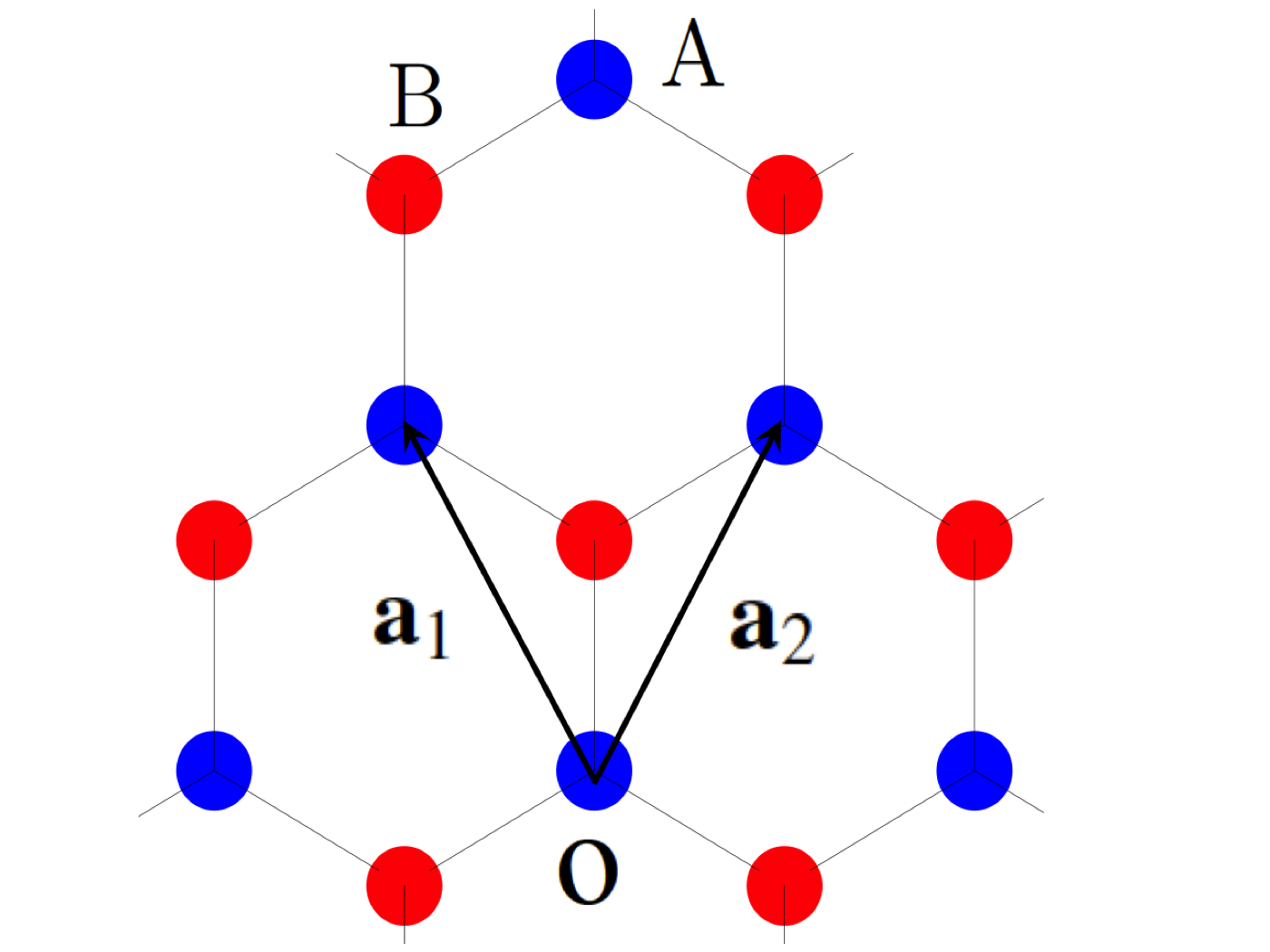} 
\label{fig:Moire(a)}}
\subfloat[]{\includegraphics[width=0.32\columnwidth , height= 0.25\columnwidth]{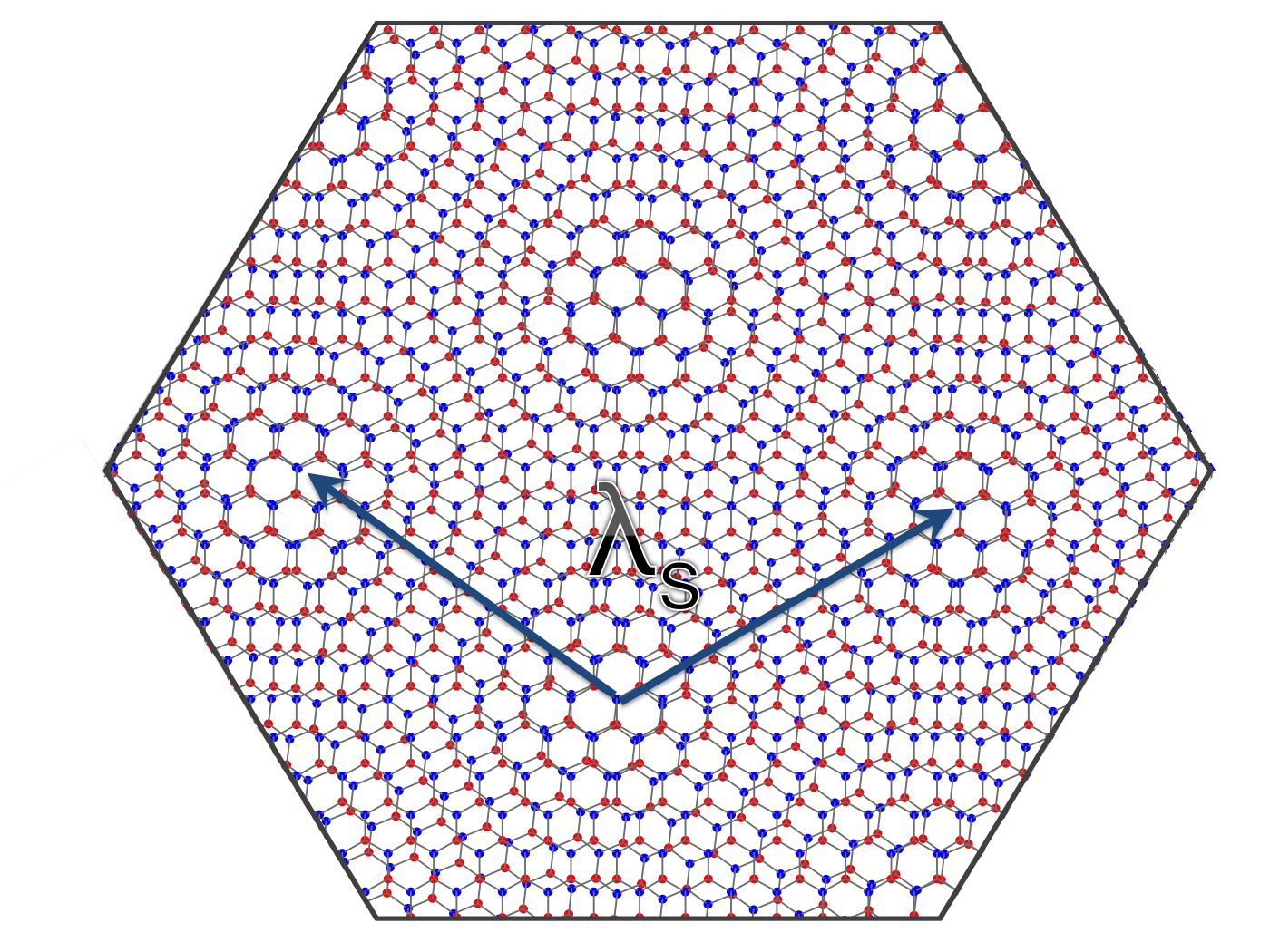} 
\label{fig:Moire(b)}}
\subfloat[]{\includegraphics[width=0.32\columnwidth , height= 0.25\columnwidth]{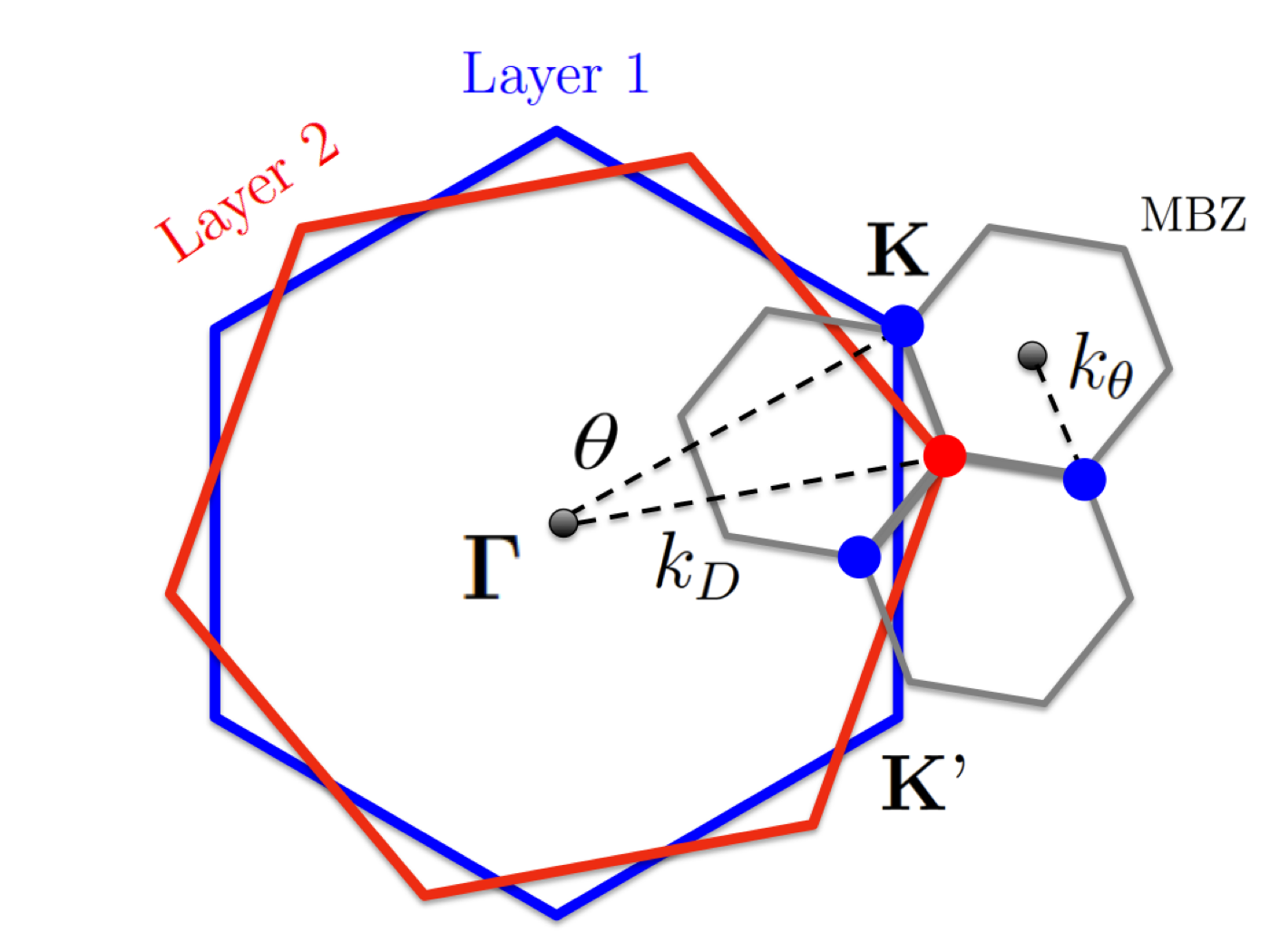} 
\label{fig:Moire(c)}}
\caption{(a) (Real space) Layer 1 of the TBLG system; (b) hexagonal (moir\'e) lattice with lattice constant $\lambda_s$ formed by twisting ($\sim \ang{10}$) a layer of graphene relative to another; (c) (k-space) The corresponding moir\'e or mini Brillouin (MBZ). }
\end{figure}

Recent reports have shown that bi-layer graphene with a small twist angle between the layers exhibits an insulating state at certain densities (bilayer~\cite{cao2018Mott, kim2017tunable} and trilayer~\cite{ChenTrilayer}) and superconductivity with a dome reminiscent of the copper-oxide materials~\cite{cao2018SC}.  This behavior is observed at a twist angle that is  close to the so-called (first) magic angle. Twisted bilayer systems are formed starting from an AB stacked bilayer and then rotating one of the layers (say layer 2) around an axis that passes through a certain A$_1$B$_2$ point (see Fig. \ref{fig:Moire(a)}). Here, A$_1$ represents sublattice A of layer 1, and B$_2$ represents sublattice B of layer 2. As the recent experiments focus on small twist angles ($\sim \ang{1}$), we confine our discussion entirely to this limit. The physics of large twist angle is equally interesting~\cite{Russians15, Kindermann18} and awaits further experimental study~\cite{RozhkovRev16}. This paper concerns the experimental claim that the insulating state preceding superconductivity is captured by Mott physics.  As will become clear, Mott physics, while absent at small twist angles, may exist, but at substantially larger twist angles where the sublattice density is sizeable.  

Electronic transport in twisted bi-layer graphene (TBLG) is dictated by an emergent moir\'e cell rather than the primitive unit cell of single-layer graphene. 
Periodicity of the moir\'e cell that originates from the repetition of the A$_1$B$_2$ sites  by demanding a commensuration condition~\cite{Shallcross10, Castro12} for which the moir\'e patterns form a superlattice with a periodicity $
\lambda_s = {a}/{2 \sin(\theta/2)} \approx {a}/{\theta} \,.$  Here $a$ ($= \SI{2.46}{\angstrom}$) is the lattice constant of the single layer graphene.  As illustrated in Fig. \ref{fig:Moire(b)}, the new cells form an effective triangular lattice with an area of $A_s = \sqrt{3} \lambda^2_s/2$ which encloses $N = (2 \lambda_s/a)^2 \sim 13000$ carbon atoms for $\theta \sim \ang{1}$. At the densities probed experimentally, only the two lowest bands in the moir\'e cell are relevant.  Because 4 electrons can occupy these bands, we define the moir\'e superlattice density
\beq
n_s =\frac{4}{A_s}=  \frac{32 \sin^2(\theta/2)}{\sqrt{3} \, a^2}  \simeq  (\theta^{\circ} )^2  \, 2. 32 \times 10^{16} \,\text{e}^{-} \,/ \text{m}^2 .
\label{eq:Ns}
\eeq

All the experimental phenomena can be indexed by the number of charges per moir\'e supercell, $\nu=n_eA_s$, where $n_e$ is the electron density.  The novel feature of these experiments~\cite{cao2018Mott, kim2017tunable}  is the presence of an insulating state at  $\nu=2$ and a weaker insulator at $\nu=3$.  Distinctly absent from this sequence is an insulating state at $\nu=1$ for which no explanation theoretical or experimental has been preferred.  Because $\nu=4$ is a band insulator, Mott physics has been invoked~\cite{cao2018Mott,balents,balents2,balents3,balents4,baskar} to explain $\nu=2,3$. However, Mott physics dictates that an insulator should exist for any filling satisfying $\nu<4$.   Additional features which point to physics beyond Mott are 1) extreme sensitivity of the insulating state to an applied magnetic field in contrast to typical Mott systems~\cite{MottDavis} and 2) a gap ($0.31\,$meV) equal to the temperature ($4\,$K) at which the insulating state obtains, unlike $\mathrm{VO}_2$, where a gap of $0.4\,$eV opens below $295\,$K.   In fact, it was always the disconnect between the temperature scale and the charge gap that Mott~\cite{mottvo2} used as evidence for the irrelevance of the structural transition to the insulating state.  In addition, in device D1~\cite{cao2018Mott}, the insulator on the electron-doped side resides at a density greater than the hole-doped density, thereby offering evidence for insulating behavior at non-integer fillings.

\begin{figure*}[htp]
\centering
\includegraphics[width=0.72\textwidth]{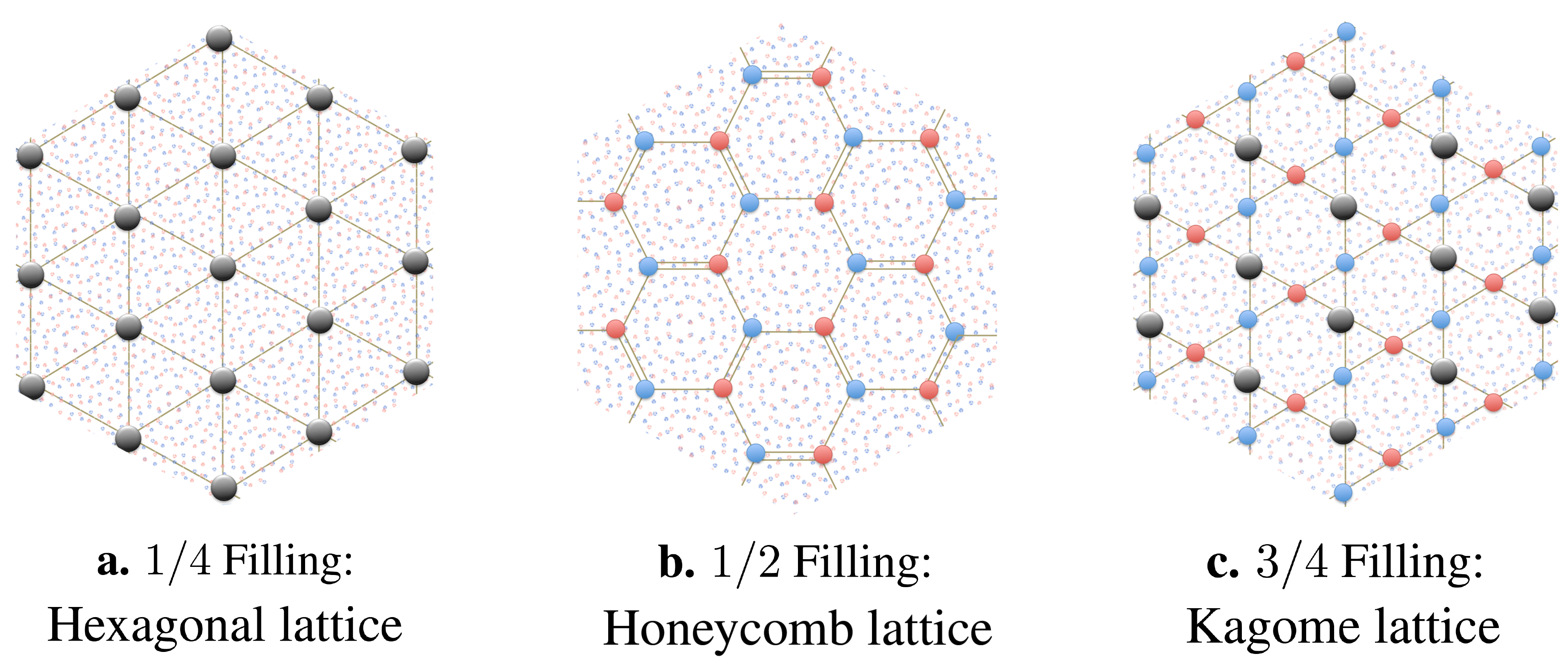} 
\caption{Proposed ground state orderings of the electron crystal for $\nu = 1, 2, 3$ electrons per unit supercell. The background is formed by the triangular moir\'e superlattice and, for clarity, the  red, blue, black dots represent the electrons localized on the supercells. The lattice constants are $\lambda_s, \lambda_s/\sqrt{3}, \lambda_s/2$ for $\nu = 1,2,3$, respectively. For $\nu=2$ the electrons can form singlets (double lines) along alternating links (in two possible ways) forming a valence bond solid, or a Kekul\'e lattice.}
\label{wclattices} 
\end{figure*}

Given these difficulties, we propose that the insulating states in TBLG are actually crystalline states on a honeycomb lattice, $\nu=2$, and a kagome lattice, $\nu=3$, see Fig. \ref{wclattices}. To establish this, we compute $r_s=E_U/E_K$ as a function of twist angle and doping. Using the full band structure, we find that $r_s$ is a non-monotonic function of the electron density and twist angle.  For $\nu=1$ the theoretical threshold of $r_s=37$ for crystallization on a triangular lattice~\cite{David37}  is never satisfied experimentally.  As a consequence, the system remains metallic.  Insulating behavior obtains for $\nu=2$ and $\nu=3$ as a result of a transition to honeycomb and kagome Wigner crystalline states.  

For a system in which the kinetic energy is of the standard quadratic form $p^2/2m$, $E_K\propto 1/r_e^2$ where $\pi r_e^2n_e=1$.  Because the potential energy scales as $1/r_e$, the dimensionless measure of the strength of the interactions scales as $r_s\propto 1/\sqrt{n_e}$.  For single-layer graphene where the dispersion is of the linear Dirac form, $r_s$ is independent of density and hence no new interaction-driven physics is expected as a function of density. It is for this reason~\cite{NoWigner,E-field} that Wigner crystallization has been argued to be non-existent in single-layer graphene (SLG). Near the magic angles, the dispersion of TBLG deviates substantially from a Dirac-like spectrum and becomes flatter with proximity to the magic angle. As a result, $r_s$ acquires dependences on both density and twist angle.  This also renormalizes  the SLG Fermi velocity, $v_0 \sim 10^6$ m/s, to an effective velocity $v_F$. For small twist angles and for an interlayer $\pi-$hybridization of $2w$~\cite{MCDMoire},
\beq
\frac{v_0}{v_F} \equiv \tilde{v}(\theta)  =  \frac{\partial_{\bs k} \varepsilon_{\bs k} |_{\bs k = \Gamma}}{\partial_{\bs k} \varepsilon_{\bs k} |_{\bs k = K, K'}} = \frac{1+6 \kappa^2}{1-3\kappa^2}\, ,
\label{eq:vtilde}
\eeq
which depends only on the ratio, $\kappa\equiv w/\hbar v_0 k_\theta \sim \mathcal{O}(1)$.  Here $k_\theta = 2 k_D \sin(\theta/2)$ is the distance of the MBZ corner (or Dirac point) to the MBZ center and $k_D = 4\pi/3 a$ is a similar measure for the larger Brillouin zone (Fig. \ref{fig:Moire(c)}).
The scope of the above formula is, of course, limited to small twist angles only and numerical methods must be used to determine this factor accurately; however, for our purposes it should be sufficient to continue with this formula\footnote{Device D1~\cite{cao2018Mott} seem to have $\tilde{v}(\ang{1.08}) = 25$, however the above formula estimates it to be about 50. We believe the origin of this discrepancy is tied to the error margin in the measurement of the twist angle itself, which is about \ang{0.1}. This introduces an error of $\pm \,40$ in the estimation of $\tilde{v}$.}. Clearly, the fermi velocity vanishes (or, flat bands occur) for $\kappa 
=1/\sqrt{3}$. The angle for which this happens is defined as the magic angle, $\theta_{\text{magic}}$ ($=\ang{1.05}$~\cite{cao2018Mott} or $\kappa \, \theta \,\, (\text{in deg.}) = 0.6 $). As we will see, availability of such flat bands will play a  pivotal role in favoring formation of WC in TBLG.

The starting point for computing $r_s$ is a 2-band, low-energy description of the system near (flat bands) the Dirac points~\cite{LowEnergyHam, KoreanEffHam, Castro11EffHam, cao2018Mott},
\beq
H_\theta (\bs k) = \bem 
0 & v_F \bs k^\dag + \frac{1}{2 m} \bs k^2 \\
v_F \bs k + \frac{1}{2 m} (\bs k^\dag)^2  & 0
\eem \, .
\label{Hamil}
\eeq 
Here $\bs k = k_x + i k_y$ and $\bs k^\dag$ denotes the complex conjugate. Note the angle dependence of the system enters through the parameters $v_F$ and band mass $m$. Diagonalizing this Hamiltonian we obtain the dispersion,
\beq
\varepsilon_{\bs k}^2 =  \left( k v_F \right)^2  +  \frac{v_F}{m} k_x \left( k_x^2 - 3 k_y^2 \right) + \left( \frac{k^2}{2m} \right)^2  \, .
\label{dispersion}
\eeq
As is evident, the band structure lacks $k_x\rightarrow k_y$ symmetry and any transport is necessarily anisotropic. This anisotropy arises strictly from the chirality of the bands.  When the system is close to the magic angle, or an almost flat band has formed ($v_F \rightarrow 0$), it is the quadratic dispersion above that becomes the leading term. Away from the magic angle, the linear term dominates. Thus, for simplicity, let us consider this dispersion order by order. For a dispersion of the form $\varepsilon_{\bs k} \sim c_n k^n$, the corresponding $r_s^{(n)}$ in 2D is given by , 
\begin{align}
r^{(n)}_s = \frac{E_U}{E_K} = \frac{e^2}{\epsilon \, c_n} \, r_e^{n-1} \, 
\label{rsn}
\end{align}
Here $\epsilon$ is the dielectric constant, which we fix to $10$ for our discussion. We justify this  value in Sec. A of Appendix Numerically it has been shown~\cite{David37} that in 2D, a WC with a triangular lattice structure (see Fig. \ref{wclattices}) occurs only for $r_s>37$.  So our goal here is twofold -- to check how close the experiments are to this threshold and to determine in which regime the Mott criterion is satisfied.  Following~\cite{cao2018Mott}, let us first consider the effects of the linear term on $r_s$.  It can be seen from Eq. \eqref{rsn}, similar to the case of SLG ($c_1 = v_F$), that 
\beq
r_s^{(1)} (\theta) =\frac{ \alpha}{ \epsilon } \, \tilde{v}(\theta) \quad , \quad   \alpha = \frac{e^2}{\hbar v_0 } \sim 2  \, 
\label{rsval1}
\eeq 
and hence is determined solely by the twist angle and not the density.  
Here $\alpha$ is the effective fine structure constant for SLG. In this system, $v_F$ can be reduced by straining~\cite{strain} or by doping~\cite{NovoDope}. However, in either case, $v_F$ never drops by more than an order of magnitude, limiting the value of $r_s$ to be no more than $2$~\cite{NoWigner}.

\begin{figure}[htp]
\centering
\includegraphics[width=0.95\columnwidth]{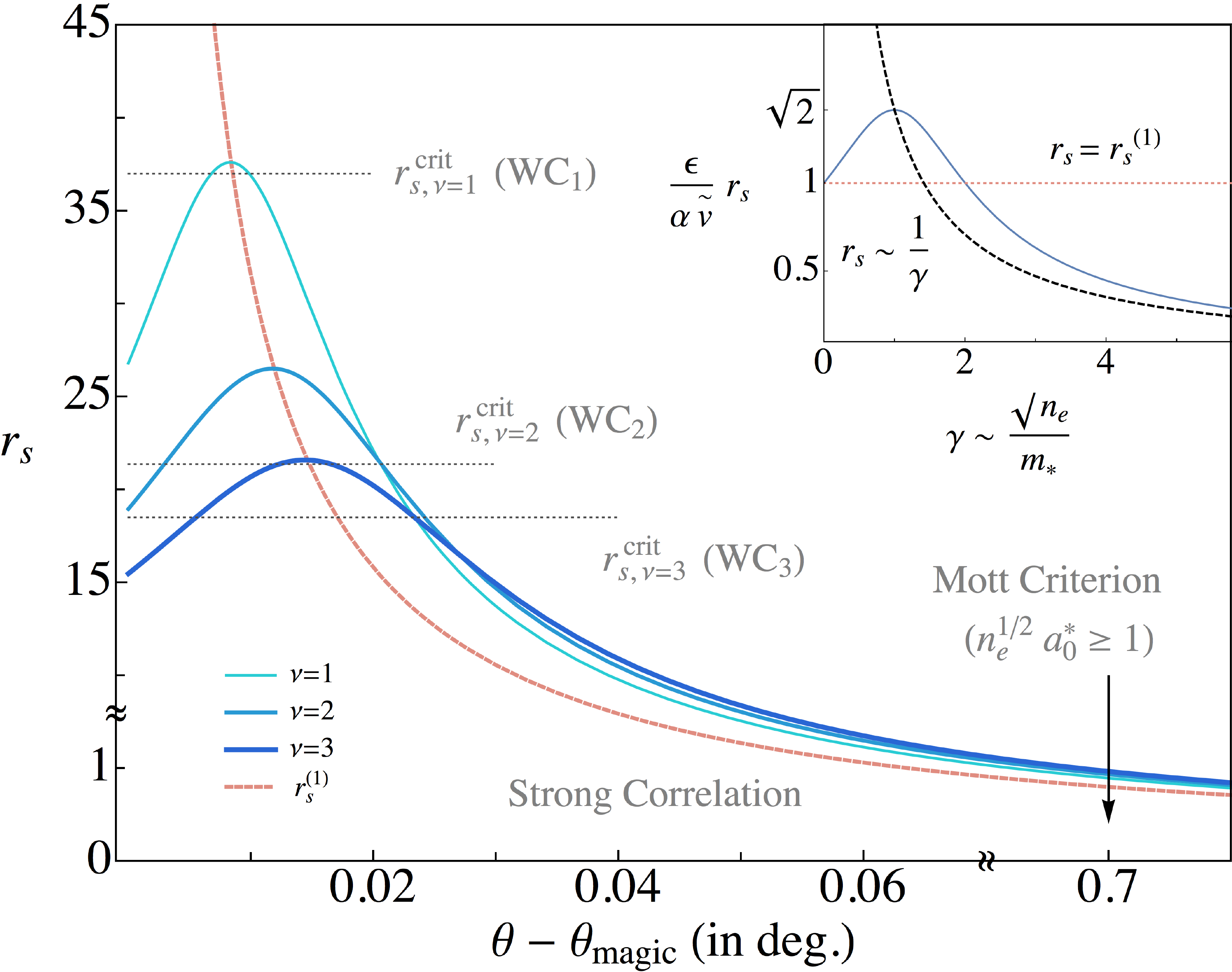} 
\caption{ The body of the figure contains the twist-angle dependence of $r_s$ [from Eq.  \eqref{rsvalnew}, with $m_\ast = 1.1 m_e$]. The blue curves are increasingly darker (or thicker) with increasing filling fraction. The dotted lines denote the critical $r_s$ required to form the WCs for a given $\nu$, $r_{s, \nu}^{\,\text{crit}} = 37, 37/\sqrt{3}, 37/2$ for $\nu = 1,2,3$, respectively.
Inset: A comparison of the $\gamma$--dependence of $r_s$ in Eq. \eqref{rsvalnew} (blue solid line) and in Eq. \eqref{rsAsym} (black dashed line). For $\gamma=0$ (orange dotted line) one recovers the expression for $r_s^{(1)}$ in Eq. \eqref{rsval1}.
}
\label{fig:Phases}
\end{figure}

TBLG requires going beyond the linear truncation used for the band structure in Eq. \eqref{rsval1}.   To this end, we consider the full low-energy dispersion of Eq. \eqref{dispersion}.  Approximating $\sqrt{2} k_x = \sqrt{2} k_y = k \sim 1/r_e$, we obtain 
\begin{gather}
r_s(\theta, \nu) = \frac{\alpha}{\epsilon} \tilde{v}(\theta) \left( 1 - \gamma +  \frac{1}{2} \gamma^2 \right)^{-1/2} \, \nn , \\
\gamma = \frac{\sqrt{2} \hbar}{ m_\ast v_F} \frac{1}{2r_e} = \frac{\theta}{a} \frac{\hbar}{m_\ast v_F} \sqrt{\frac{ \pi \nu}{\sqrt{3}}} \sim \frac{\theta^{\,\circ}}{100} \, \frac{\tilde{v}(\theta) }{ \tilde{m}(\theta, \nu)} \sqrt{\nu}  \, ,
\label{rsvalnew}
\end{gather}
which should be valid arbitrarily close to the magic angle. Here we have defined $m_\ast/ m_e \equiv \tilde{m}$. The $-\gamma$ term enters because of the chiral nature (odd powers of the $k_x+ik_y$ term) of the Dirac bands and ultimately gives rise to non-monotonicity of $r_s$. 
A physical significance of  the dimensionless density parameter $\gamma$ ($\sim \sqrt{n_e}/m_\ast$, for a fixed $\theta$) is that it measures the wavelength of the (renormalized) electrons in units of the superlattice constant (the bare electron wavelength is $\sim 0.6\,a$, which fixes the factor $\pi/300 \approx 1/100$ after converting $\theta$ into degrees).  Noting that $\gamma=1$ maximizes Eq. \eqref{rsvalnew}, we plot $r_s$ as a function of the twist angle and $\gamma$ in Fig. \ref{fig:Phases}. There are two key features. First, notice with increasing $\gamma$ or density,  $r_s$ increases and then it decreases. This non-monotonicity is due to the presence of the negative sign in Eq. \eqref{rsvalnew}, which originates from the fact that we are considering chiral bands (Hamiltonian is complex valued in $\bs k$).  Secondly, when $\theta \approx \theta_{\text{magic}}$, $\tilde{v} \rightarrow \infty$, causing $\gamma$ to increase rapidly and the corrections to the linear dispersion become important, restricting one to the right-end of the graph. For $\tilde{v} \rightarrow \infty$ we obtain an asymptotic expansion of Eq. \eqref{rsvalnew},
\beq
r_s(\theta_{\text{magic}}, n_e) \simeq  \frac{r_s^{(\text{max})}}{\gamma} \sim  \frac{m_\ast(n_e)}{\sqrt{|n_e|}} \,\, , \,\,  r_s^{(\text{max})} = \tilde{v} \, \sqrt{2} \, \frac{\alpha}{\epsilon}\, .
\label{rsAsym} 
\eeq

Note that both $\tilde{v}$ and $\gamma$ are diverging near the magic angle but their ratio is not. This is shown as the (orange) dashed line in Fig. \ref{fig:Phases}. We will see this feature in the sample M2 of~\cite{cao2018SC}.    To expose the twist-angle dependence of $r_s$, we plot  in Fig. \ref{fig:Phases} the computed values of $r_s$ both at the linear dispersion level (dashed orange line in Fig. \ref{fig:Phases}), Eq. \eqref{rsval1} and the full expression, Eq. \eqref{rsvalnew}.   
Unlike the result for the linear dispersion, the full treatment yields an $r_s$ that remains finite close to the magic angle.  An additional feature highlighted in Fig. \ref{fig:Phases} is the critical value of $r_s$  into the WC shown in Fig. \ref{wclattices} as a function of $\nu$.  If a WC with $\nu>1$ charges per unit cell simply amounts to replacing $e$ with $e\nu$ in the triangular lattice shown in Fig. \ref{wclattices}a for $\nu=1$, then the critical $r_s$ would simply decrease by $1/\nu^2$.  However, this assumes the formation of charge puddles, ignoring any internal structure. This can be taken care by conforming to the structures of Fig. \ref{wclattices}. Since $r_s$ scales with distance, the critical $r_s$ should be related to the ratio, $d_\nu$, of the lattice  constant in the new lattice relative to that in the triangular lattice~\cite{DConv}.   The only choice for $\nu=2$ is a honeycomb lattice~\cite{balents2} as shown in Fig. \ref{wclattices}b.  The double lines indicate singlet pairs. Resonating structures based on this kind of singlet ground state have also been proposed for $\nu=2$~\cite{LeeKekule, Subir}.  Likewise, the analogous structure for $\nu=3$ can be obtained by placing 3 charges in a moir\'e cell and leads to the kagome lattice shown in Fig. \ref{wclattices}.   Simple geometry shows that for $\nu = 1, 2, 3$, the values of $d_\nu = 1, 1/\sqrt{3}, 1/2$. Consequently, $r_{s, \nu}^{\text{crit}} = 37 d_\nu = 37, 21, 18.5$, respectively. Each of these critical values are indicated in Fig. \ref{fig:Phases}.  Our major conclusion is that for $\nu=1$, the threshold of $r_s=37$ is never achieved experimentally, while for $\nu=2$ and $\nu=3$ is easily achieved depending on the magic angle.  Consequently, this model explains the nature and occurrence of the insulating states in TBLG.   The fact that we have bounded the transition, $37/\nu^2 < r_{s, \nu}^{\text{crit}}\lesssim 37 d_\nu$, makes the arguments presented here significant because either of these limits predicts that WC formation is feasible where the insulating states are seen and metallic behavior for $\nu=1$.    The precise location of the transition will of course require an extensive quantum Monte-Carlo calculation but given the difficulty with the sign problem, the estimates above should suffice~\cite{DConv} to ground the argument presented here.

Now to rule out Mottness, we evaluate the dynamical criterion for the Mott transition~\cite{MottDavis}
\begin{align}
n_e^{1/2} a_0^\ast  \gtrsim 1.
\label{mottcriterion}
\end{align}
Here $a_0^* = {\hbar^2}/{m^{\phantom{*}}_* e_*^2} $ is the effective Bohr radius, the renormalized electronic charge is $e^2_* = e^2/ \epsilon $ and $m_\ast$ is the effective mass. In order to address this criterion we adopt two methods of obtaining the effective mass. 

\textit{Method I}-- The dispersion relation in Eq. \eqref{dispersion} corresponds to an  effective mass of
\bea
m_\ast(\theta, \nu) = m_e  \, \frac{\left(\gamma^2 - 2 \gamma + 1\right)^{3/2}}{\gamma^3 - 3 \gamma^2 + .45 \gamma - 2} \, .
\eea
Here $\gamma$ is the same as that defined in Eq. \eqref{rsvalnew}, except here we fix the mass appearing there to $m_e$. This allows us to write the Mott-criterion as 
\beq
\frac{\lambda_s(\theta)}{a_0} \, \frac{m_\ast(\theta, \nu)}{m_e \sqrt{\nu}} \lesssim \epsilon \, .
\eeq
Notice, however, near the magic angle $\gamma$ diverges, causing $m_\ast \simeq m_e$. Thus, the left-hand side of the above inequality is decided entirely by the factor $\lambda_s/a_0$. In the low-angle limit $\lambda_s/a_0 \gg 1$, typically  around $150-250$. This makes it impossible for the system to satisfy the Mott criterion, irrespective of the filling-fraction. This constitutes the most stringent test of Mottness and TBLG fails it.

\textit{Method II}. -- Instead of using the above formulae one can work with $m_\ast /m_e=  \tilde{v} \, \sqrt{{h^2 n_e }/{8 \pi v_0^2 m_e^2}}$. This is obtained using a zero-temperature density of states formula for a linear band, which also seems to provide a good fit for the quantum oscillation data near half-filling (Fig. 3b) of~\cite{cao2018Mott}. Here the factor $8$ arises from the degree of degeneracy, $\textsl{g}$, (spin, valley, layer) of the bilayer system. Using this we simplify Eq. \eqref{mottcriterion} to 
\beq
r_s^{(1)} (\theta) \lesssim \frac{1}{\sqrt{\pi}}  \, .
\label{rs1Mott}
\eeq
It is the $\sqrt{n_e}$ scaling of the effective mass that makes the above criterion density independent.  Note a heuristic formula in which $n_e^{1/2}$ is compared with $A_s^{1/2}$ contains no dynamical information and hence is not relevant to the Mott transition.  Combining Eq. \eqref{rs1Mott} with Eq. \eqref{rsval1}, we find that the Mott criterion is satisfied at $\theta_{\text{Mott}} \sim \ang{1.7}$, substantially away from the magic angle!  Hence, even beyond phenomenological concerns, from a purely theoretical standpoint,  Mottness can be ruled out for TBLG.

\begin{figure}
\centering
\includegraphics[width=0.95\columnwidth]{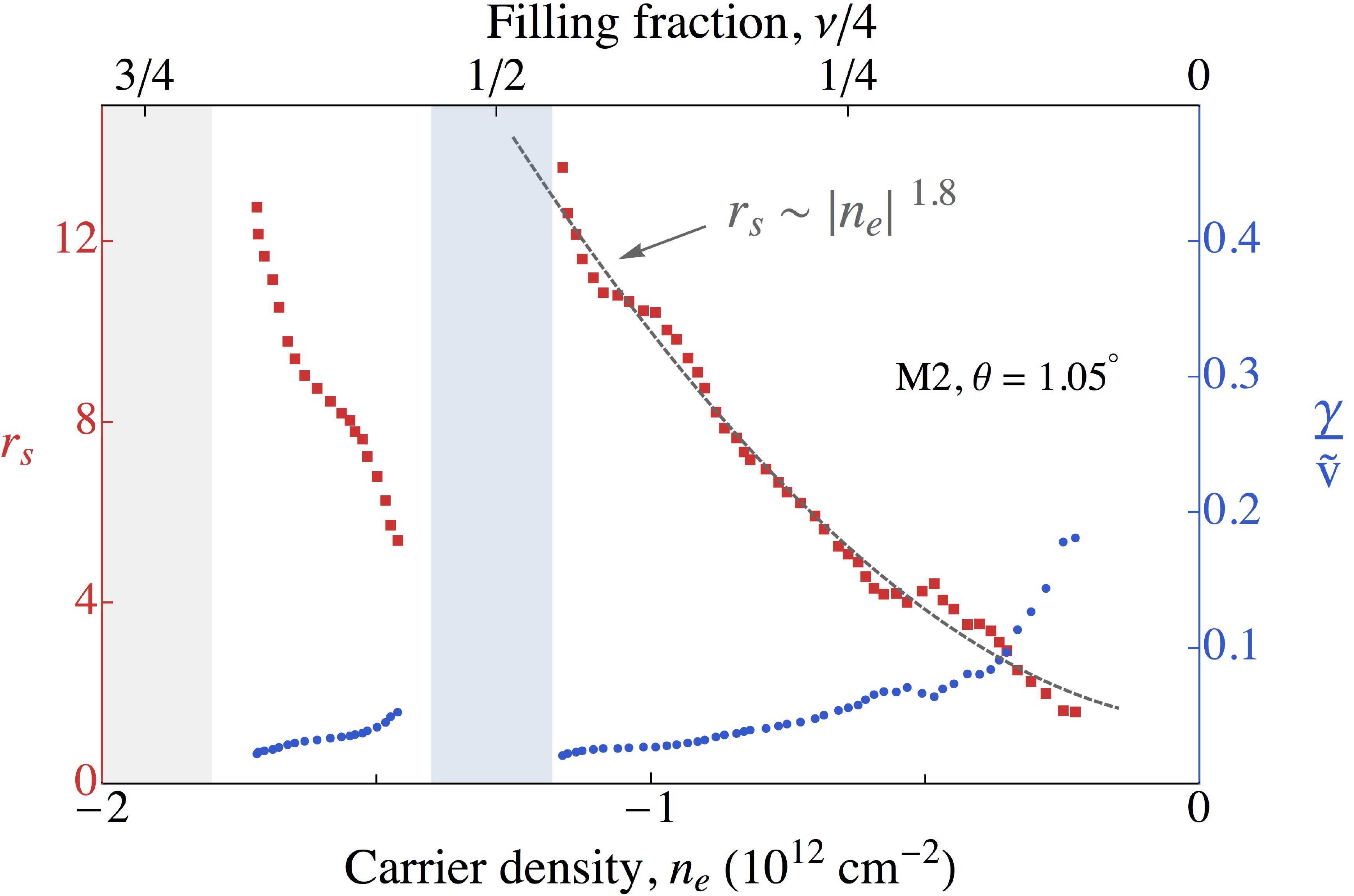} 
\caption{By using effective mass data for sample M2 [Fig. 5e of~\cite{cao2018SC}] we obtain $r_s$ (red squares) with help of the expression in Eq. \eqref{rsvalnew}. The accompanying blue dots are $\gamma$ defined there. Since $\theta \approx \theta_{\text{magic}}$, we see a behavior that can also be governed by Eq. \eqref{rsAsym}. The gray (dashed) curve is a power-law fitting of $r_s$ ($\sim |n_e|^{1.80 \pm 0.2} $) for low doping until half-filling. 
}
\label{fig:rsM2}
\end{figure}

Strictly for calibration purposes, we can apply these results immediately to the experiments.  To this end, we extract $m_\ast$ from the experimental data and compute the resultant $r_s$ using Eq. \eqref{rsvalnew} and consider devices D1 and M2.  The results for D1 are in Sec. B of Appendix. Recall M2 is a superconducting sample with $\theta = \ang{1.05}$. The results are shown in Fig. \ref{fig:rsM2}. Clearly in this case the $r_s$ does not show any dome feature, which can be understood from (the inset of) Fig. \ref{fig:Phases}. In other words, extreme proximity to the magic angle enhances $\gamma$, thereby causing $r_s$ to show roughly  monotonic behavior with density, as approximated by Eq. \eqref{rsAsym}.   It may seem odd that with increasing density $r_s$ increases.   This state of affairs obtains  because the effective mass is non-trivially dependent on doping and is able to grow faster (empirically fitted with $m_\ast \sim n_e^{2.3}$) than $\sqrt{n}$ causing $r_s$ to increase.  In M2, the dip in the conductance at $\nu= 2$ mirrors that at $\nu=-2$ and occurs roughly where $r_s$ is maximized.  Hence, in this case, the insulating state does arise from an enhancement in the correlations $\nu=\pm 2$.   We were unable to compute $r_s$ all the way to $\nu=2$ because of the lack of experimental data for $m^\ast$ in this regime.  Nonetheless, the behavior of $r_s$ is consistent with the reduced critical values of $r_s$ needed to crossover to the lattices shown in Fig. \ref{wclattices} for $\nu=2$ and $\nu=3$ while such a transition is not feasible for $\nu=1$. 

As a result, our major claim based on a computation of $\theta_{\text{Wigner}}$ and $\theta_{\text{Mott}}$ is that the experiments fall into the Wigner not the Mott regime in particular with insulating states for $\nu=2$ and $\nu=3$ of the honeycomb and kagome kind.   The Lindemann criterion places the melting energy at roughly $1\%$ of the intra-cell Coulomb energy roughly 10-30meV\~cite{cao2018Mott} and hence is consistent with the temperature of $4K$ where the metallic state ensues.  Mott physics may obtain at considerably larger twist angles where the traditional Mott criterion~\cite{MottDavis} applies.  While it is unclear if such experiments can be performed, this work raises the possibility of large twist angle physics as a potentially new transport regime.
 The correct framework for analyzing superconductivity should be from doping the honeycomb and kagome lattices shown in Fig. \ref{wclattices}. It was previously argued that the retardation effects that persist for $r_s\gg 1$ can lead to superconductivity in the vicinity of the melting (doping) of a WC~\cite{phillipsNature}. Thus, the superconductivity observed in Ref.~\cite{cao2018SC} could be a direct consequence of the superconducting correlations that reside in close proximity to WC.  Hence, this work highlights the new physics that should reside in the unsolved large $r_s$ parameter space.

\textit{Acknowledgement}: We are thankful to David Ceperley, Bryan Clark, Liang Fu, Pablo Jarillo-Herrero, Tero Heikkil\"a,  Peter Silvestrov, Louk Rademaker, Antonio H. Castro-Neto for many helpful discussions and the anonymous referee for suggestions on the previous version of the paper. We acknowledge support from the Center for Emergent Superconductivity, a DOE Energy Frontier Research Center, Grant No. DE-AC0298CH1088.  We also thank the NSF DMR-1461952 for partial funding of this project.

\textit{Note added}: During the revision of this manuscript, we became aware of Ref.~\cite{TeroMFT} which finds failure of mean-filed theory to explain the correlated insulator. This supports our conclusion of large $r_s$ in TBLG. We also became aware of some recent works where a non-Mott picture is argued, such as Kekul\'e valence bond order~\cite{LeeKekule}, quantum spin liquid~\cite{SpinLiq}, charge-density wave order~\cite{LiangFu}, chiral spin-density wave order~\cite{FanYang}, Dirac semimetal state~\cite{Ochi}, long-range antiferromagnetic order~\cite{Subir} and nematic phase~\cite{Kivelson}.
%

%
%
%
%
%
\onecolumngrid 
\renewcommand{\thefigure}{A\arabic{figure}}
\setcounter{figure}{0}
\renewcommand{\theequation}{A\arabic{equation}}
\setcounter{equation}{0}

\section{Appendix A: Effects of Screening} 
\label{app:Screening}

The calculation presented in the main text assumes the presence of a long-range Coulomb interaction. However, this might be a bit stringent assumption especially for a system consisting of graphene layers. In order to understand the corrections originating from screening we evaluate the dielectric function under the random phase approximation (RPA). In particular, in this Appendix we compute the RPA corrected dielectric function under the static limit ($\omega = 0$),
\bea
\epsilon_{\text{RPA}}(\bs q) = \epsilon \left[ 1 - v_q \Pi(\bs q) \right] \,.
\label{eRPA}
\eea
Here $v_q = 2 \pi e^2/ \epsilon q$ is the unscreened 2D Coulomb interaction, and $\Pi(\bs q)$ is the single-particle bubble. Since we will be limiting our discussion to $T \rightarrow 0$, we express $\Pi(\bs q)$ only in terms of the inter-band scattering term,
\bea
\Pi(\bs q) = - \textsl{g} \int \, \frac{d^2 k}{(2 \pi)^2} \frac{2}{\varepsilon_{\bs k} + \varepsilon_{\bs k'}} \, \left \rvert  \psi_+^\dag (\bs k) \psi_- (\bs k') \right \rvert^2 \, .
\eea
Here, $2$ appears since we work with a 2-band system and  $\varepsilon_{\bs k}$ is the eigenvalue of the low-energy Hamiltonian in Eq. \eqref{Hamil} [or see Eq. \eqref{dispersion}] and the eigenvectors of this Hamiltonian are
\bea
\psi_+(\bs k)  = \frac{1}{\sqrt{2} \varepsilon_{\bs k}}  
\bem v_F \bs k + \frac{1}{2m} ({\bs k}^\dag)^2 \\ \varepsilon_{\bs k}  \eem  \quad , \quad 
\psi_-(\bs k)  = \frac{1}{\sqrt{2} \varepsilon_{\bs k}}  
\bem - \varepsilon_{\bs k}  \\ v_F {\bs k} + \frac{1}{2m} {\bs k}^2  \eem   \, .
\eea
Our notation here is, $\bs k = k_x + i k_y$, $\bar{\bs k} = k_x - i k_y$, $\bs k' = \bs k - \bs q$, and $\vec{k} = k (\cos \theta_{\bs k}, \sin \theta_{\bs k})$. The inter-band scattering cross-section can be simplified to
\bea
2 \varepsilon_{\bs k} \varepsilon_{\bs k'} \left \rvert  \psi_+^\dag (\bs k) \psi_-(\bs k') \right \rvert^2 &=& \varepsilon_{\bs k} \varepsilon_{\bs k'} - v_F^2  \vec{k} \cdot \vec{k}' + \frac{k^2 {k'}^2}{4m^2} \left[1 - \frac{2}{k^2 {k'}^2}  \left( \vec{k} \cdot \vec{k}' \right)^2\right] 
\nn \\ &+& 
\frac{3 k^2 v_F}{2m} \big[ q_x \cos(2 \theta_{\bs k}) - q_y \sin(2 \theta_{\bs k}) \big] \, .
\label{scatter}
\eea
The first term is simply the vacuum term, the second one is for  SLG and the third one arises for bi-layer graphene. The last term, an intermediate term in powers of momenta, is present purely due to the twist angle. In obtaining the above form of this term, we have in fact dropped several other terms since they do not contribute to the bubble due to being odd-powered in momenta.

Instead of considering the full integration, which anyway is daunting task to perform analytically, we consider only the relevant limits. In fact, for $r_s \gtrsim 1$ the validity of such a perturbative expression in Eq. \eqref{eRPA} is questionable. In other words, for twist angles close to magic angle ($\theta - \theta_{\text{magic}} \lesssim  \ang{0.5}$) such a calculation breaks down. We thus limit our discussion on screening-effects only to twist angles away from the magic angle, in other words when the dispersion is predominantly linear. In this case, only the first two terms in Eq. \eqref{scatter} could contribute, 
\beq
\text{SLG Limit}: \hspace{1cm} E_U = \frac{1}{ \epsilon_{\text{RPA}}} \frac{e^2}{r} \quad , \quad  \epsilon_{\text{RPA}} = \epsilon \left( 1 + \frac{\pi}{8}\textsl{g}\, r_s^{(1)} \right) = \epsilon + \pi \alpha \, \tilde{v}(\theta) \, .
\eeq
Here $\epsilon = 3-4$ is the dielectric constant of the hBN substrate~\cite{hBNchem} used for the TBLG sample and recall $\textsl{g}=8$. This result can be understood by the following argument, since $\varepsilon_{\bs k} \sim k$ amounts to the high-energy limit, the electrons of the two layers are almost decoupled, leading to a long-range and sing-layer like behavior, as obtained in Ref.~\cite{HwangSLG} for SLG. This corrects the value of $r_s^{(1)}$ to 
\beq
r_s^{(1, \text{ RPA})} = \frac{ r_s^{(1)}}{1 + \pi r_s^{(1)}} \lesssim \frac1\pi \, .
\eeq
This is clearly much less than 1. In fact, for all practical twist angle considerations, one can assume $r_s$ to be $1/\pi$, independent of the angle. However, we will be mostly interested in angles closer to the magic angle, for which we continue using $r_s(\theta, \nu)$ of Eq. \eqref{rsvalnew}. Keeping this in mind, we simply set $\tilde{v} =1$ and work with $\epsilon_{\text{RPA}} = 4 + \pi \alpha \approx 10$, that is precisely the value used in Ref.~\cite{cao2018Mott}.

Although not very meaningful, for the sake of completeness, we consider the other limit when the twist angle is very close to magic angle and thus making the (1st and) 3rd term in Eq. \eqref{scatter} dominate. In this limit~\cite{SDSGap}, $\Pi(\bs q) = - N_0 \log 4$, where $N_0 = \textsl{g} m/2 \pi $ is the density of states in the bi-layer graphene  limit. The dielectric function becomes, $\epsilon_{\text{RPA}}/\epsilon = 1 + q_{\text{TF}}/q$, where $q_{\text{TF}} =  ( \log 4) \textsl{g} \,m e^2/ \epsilon $ is the Thomas-Fermi screening vector. This screens the Coulomb interaction to 
\bea
\text{BLG Limit}: \hspace{1.5cm} E_U = \frac{2 \pi e^2}{\epsilon \left(q + q_{\text{TF}}\right)} \xrightarrow{r \gg 1/q_{\text{TF}}}\frac{1}{\epsilon q_\text{TF}^2}  \, \frac{e^2}{r^3} .
\eea

\renewcommand{\thefigure}{B\arabic{figure}}
\setcounter{figure}{0}
\section{Appendix B: Device D1 Modeling}
\label{app:D1}

\begin{figure}
\centering
\includegraphics[width=0.85\columnwidth]{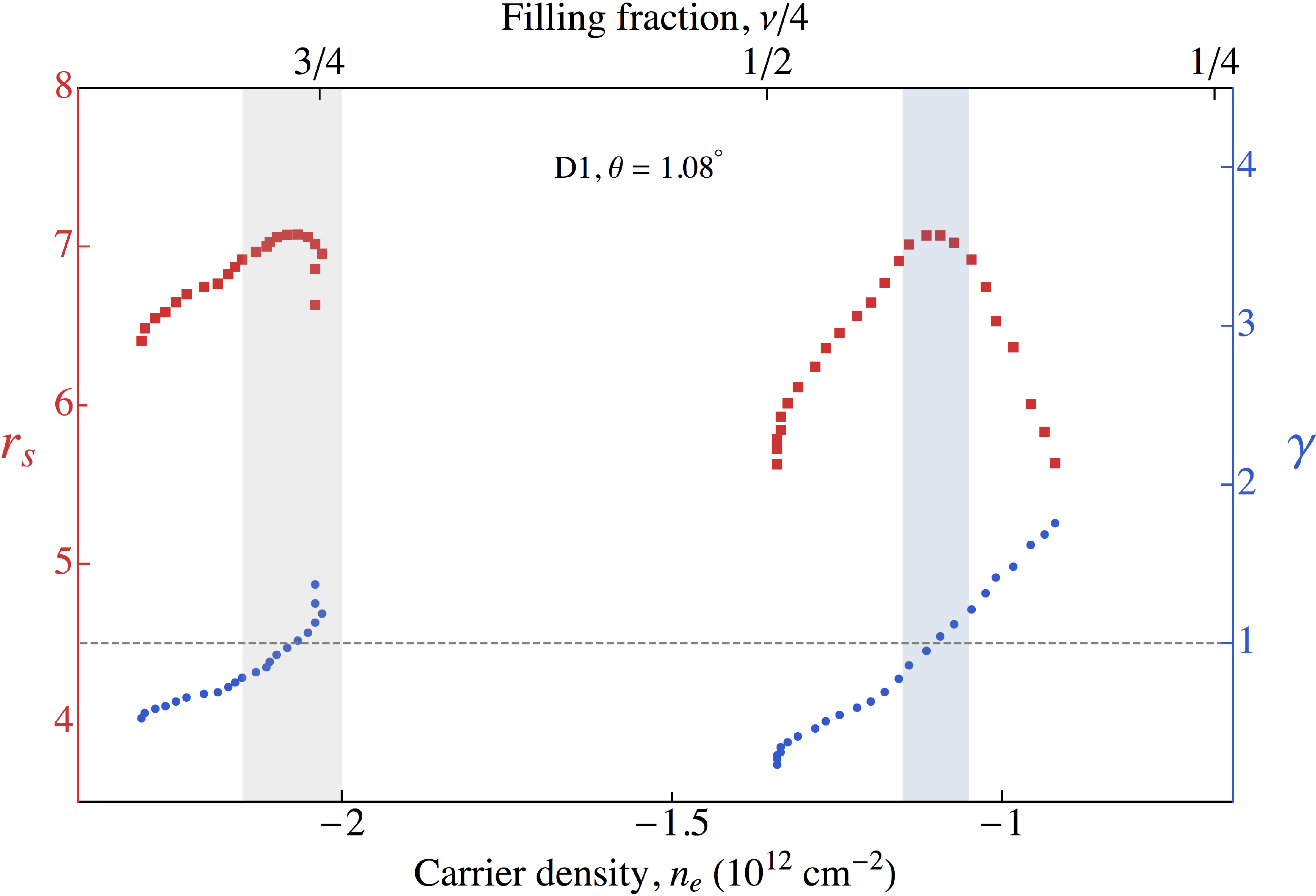}
\caption{We redraw Fig. \ref{fig:rsM2} of the main text, by using the effective mass data for sample D1. We observe maximization of $r_s$ slightly away from $1/2$-filling but close to $3/4$-filling. The dome-like feature is a reminiscent of the one present in Fig. \ref{fig:Phases} (inset) in the main text. Since D1 is slightly away from the magic angle, it can access the small-$\gamma$ region and achieve this maximization.}
\label{fig:rsD1}
\end{figure}

Here we apply the same method we used for M2 but for device D1.  For that we extract $m_\ast$ as a function of carrier density (Fig. 3b of~\cite{cao2018Mott}) and compute the resultant $r_s$ using Eq. \eqref{rsvalnew}.  Recall, D1 contains the discrepancy of $\nu=2$ for hole-doped half-filling and $\nu=2.2$ for electron doping.  We obtain the value of $r_s$ using Eq. \eqref{rsvalnew} which is shown in Fig. \ref{fig:rsD1}.  We clearly see the dome-like behavior of $r_s$ near $1/2$ and $3/4$ filling, a direct result of the dome seen in Fig. \ref{fig:Phases} near $\gamma=1$.  As in the experimental data, there is no feature in $r_s$ at $\nu=1$.  Also note the displacement of the peak in $r_s$ away from  $\nu=2$. The dip in the conductance in Ref.~\cite{cao2018Mott} occurs also occurs away from $\nu=2$, however, in the opposite direction. Hence, at present no conclusion can be made.   

\end{document}